\documentclass[10pt,journal,compsoc]{IEEEtran}
\usepackage{blindtext}
\usepackage{hyperref}

\usepackage{amsmath,amssymb,amsfonts}
\usepackage{hyperref}
\usepackage{enumitem}
\usepackage{algorithmic}
\usepackage{graphicx}
\usepackage{textcomp}
\usepackage{supertabular}

\ifCLASSOPTIONcompsoc
   \usepackage[nocompress]{cite}
\else
    \usepackage{cite}
\fi

\ifCLASSINFOpdf
 
\else
 
\fi

\usepackage{graphicx}
\usepackage{xcolor}

\begin{document}

\title{A systematic literature review of cyberwarfare and state-sponsored hacking teams}

\author{\IEEEauthorblockN{ Darshan Harsora\IEEEauthorrefmark{1}  }\\
\IEEEauthorblockA{\textit{\IEEEauthorrefmark{1}School of Engineering,} \\
\textit{University of Guelph, Ontario, Canada  }\\
dharsora@uoguelph.ca}
\\
 \and
\IEEEauthorblockN{ Khushalkumar Khoyani\IEEEauthorrefmark{2}}\\
\IEEEauthorblockA{\textit{\IEEEauthorrefmark{2}School of Engineering,} \\
\textit{University of Guelph, Ontario, Canada }\\
kkhoyani@uoguelph.ca}}

%\author{Michael~Shell,~\IEEEmembership{Member,~IEEE,}
        %John~Doe,~\IEEEmembership{Fellow,~OSA,}
        %and~Jane~Doe,~\IEEEmembership{Life~Fellow,~IEEE}% <-this % stops a space
%\IEEEcompsocitemizethanks{\IEEEcompsocthanksitem M. Shell was with the Department
%of Electrical and Computer Engineering, Georgia Institute of Technology, Atlanta,
%GA, 30332.\protect\\

%E-mail: see http://www.michaelshell.org/contact.html
%\IEEEcompsocthanksitem J. Doe and J. Doe are with Anonymous University.}% <-this % stops an unwanted space
%\thanks{Manuscript received April 19, 2005; revised August 26, 2015.}}

%\markboth{Journal of \LaTeX\ Class Files,~Vol.~14, No.~8, August~2015}%
%{Shell \MakeLowercase{\textit{et al.}}: Bare Demo of IEEEtran.cls for Computer Society Journals}

\IEEEtitleabstractindextext{%
\section*{Abstract}
\textcolor{black}{{It is expected that the creation of next-generation wireless networks would result in the availability of high-speed and low-latency connectivity for every part of our life. As a result, it is important that the network is secure. The network's security environment has grown more complicated as a result of the growing number of devices and the diversity of services that 5G will provide. This is why it is important that the development of effective security solutions is carried out early. Our findings of this review have revealed the various directions that will be pursued in the development of next-generation wireless networks. Some of these include the use of Artificial Intelligence and Software Defined Mobile Networks. The threat environment for 5G networks, security weaknesses in the new technology paradigms that 5G will embrace, and provided solutions presented in the key studies in the field of 5G cyber security are all described in this systematic literature review for prospective researchers. Future research directions to protect wireless networks beyond 5G are also covered.}}

% Note that keywords are not normally used for peerreview papers.
\begin{IEEEkeywords}
cyberwarfare, cyber forces, hacked, cyber security, Information warfare, cyberattack.
\end{IEEEkeywords}}

% make the title area
\maketitle

\IEEEdisplaynontitleabstractindextext

\IEEEpeerreviewmaketitle

\section{Introduction}\label{sec:introduction}
Cyberspace has grown with the Internet's quick expansion. The growth of cyberspace is significantly altering and influencing communication practices. Cyberspace, after the land, sea, air, and space battlefields, has become more important in the military industry. Cyberattacks of all kinds are used to advance and produce cyberwarfare in cyberspace \cite{paper1}.In this type of cyber risk environment, it's critical to swiftly recognize and recognize specific attack information in order to successfully fight against assaults. The current cyber defense system, however, is concentrated on repair after damage has been done. The defensive system that is solely concerned with recuperation has a drawback since it is hard to react quickly in between real operations. The recovery point defense system's drawback is that it is hard to react quickly in between operational phases. A new framework for cyberwarfare operations is required to support this. Every nation develops its hacking teams in a unique way. After the army, navy, and air force, there is a distinct cyber force, which is the fourth force. However, many of them are informal since they take the shape of task forces made up of different community members who have been assigned particular duties. This cyber army is made up of a variety of people with exceptional and distinctive skills who work together to create a resilience capacity against new cyberattacks \cite{paper2}.

There are some peculiar aspects of cyber warfare. Since the invention of nuclear weapons and intercontinental missiles, it is perhaps the first significant new style of warfare. Due to its novelty, there is now a virtual policy space there are no educated, transparent, political, or public conversations about what would constitute an ethical and prudent approach to the employment of such weapons. The "attribution problem" refers to the difficulty in identifying the source of cyberattacks. Since many nations can credibly assert that the cyberattacks may have begun from within their borders but that their governments did not start them, this fact would provide many cyberattacks credible immunity from prosecution \cite{paper3}.

\subsection{Prior research}
Regarding the cyberwarfare and hacking subject of our review, we have located a sizable number of publications, research papers, and periodicals. The information on this subject is still scant and insufficient. According to a study on cyberwarfare by Richardus Eko Indrajit, Marsetio, Rudy Gultom, and Pujo Widodo \cite{9631306}, the cyber army engage in four different types of operations. Each pertains to attempts to: I defend important national assets; (ii) cope with incoming assaults; (iii) launch attacks to render the opposing side helpless; and (iv) wage war on different enemies. Additionally, it is done to develop the knowledge and abilities required to be able to execute these four activities.The findings indicate that thirty professions predominate in the building of the collective cyber army capability.

"An operational website whose distinct and distinctive characteristics are framed by the use of devices and the emission spectra to create, store, modify, exchange, and manipulate information through interconnected digitalization (ICT) based systems and their associated infrastructures," according to Dan Kuehl\cite{paper5}. This fits with how we initially defined the cyberworld: as any virtual reality confined within a group of devices and networks. There are several cyberworlds, but the Internet is the one that has the most bearing on the issue of cyberwarfare right now. Cyberwarfare combines assault, defense, and special technical activities on computer networks \cite{shelton1998joint}.

Government websites in Ukraine were hacked, taken over, and flooded with threatening comments during January 14–16, 2022. A report detailing the discovery of "destructive cyberattacks aimed directly against Ukraine's digitalization" was published online by Microsoft's Threat Information Centre (MSTIC) on January 15; days earlier, attackers were using a Structured Query Language altitude weakness to insert "wiper" ransomware in Ukrainian servers. As a result, several Ukrainian organizations had their data wiped away on January 18\cite{paper7}.

\subsection{Research goals}
The goals of this study are the analysis of previous research and its conclusions, as well as a summary of the efforts of cyberwarfare research. As stated below in table \ref{Mytable1}, we created three research questions to help us concentrate the effort.
\begin{table}[h!]
\caption{Research questions.} 
\label{Mytable1}
\setlength{\tabcolsep}{3pt}
\begin{tabular}{p{106pt} p{109pt} }
\hline
 Research Questions(RQ) & Discussion\\
 \hline
 \vspace{0.15 mm} \textbf{RQ1:} What is cyberwarfare? 
  & \vspace{0.15 mm}Cyberwarfare includes nation-state activity.or a foreign government or group to engage in harm the information or computers of another country networks, such as by computer viruses or assaults that disrupt service.
\\
 \vspace{0.15 mm}  \textbf{RQ2:} Is cyberwarfare ethical?
 & \vspace{0.15 mm}To assess the ethically appropriate military responses ethical and legal in the face of a cyber-attack Norms focused primarily on what is prohibited Physical confrontations are less obvious than relevant to cyberattacks without taking into account the target has a serious, damaging physical potential irrelevant code \\
 \vspace{0.15 mm}  \textbf{RQ3:} What kind of infrastructure is required to cope with cyberwarfare?
 & \vspace{0.15 mm} utilities must create effective incident response systems. discuss their network's best practices and make strategies, and promote openness by alerting authorities to a governmental body. a number of nations have established cyber soldiers or armies whose mission is the protection the nation from disasters brought on by online assaults.\\
 \hline
\end{tabular}
\end{table}
\subsection{Contribution and Layouts}
For anyone wishing to advance their work in cyberwarfare and cybersecurity, this SLR offers the following contributions:
\begin{itemize}
\item We find 61 primary studies on hacking and cyberwarfare.Other academics might utilize this collection of papers to deepen their understanding of this topic.
\item We further choose 15 primary research that satisfy the standards we establish for quality evaluation. These papers can serve as appropriate comparison points for examination of related research.
\item To communicate the concepts and problems in the disciplines of science of cyberwarfare, hacking, and cyber security, we undertake an extensive examination of the data included in the subset of 15 papers.
\item We give a meta-analysis of such current situation with relation to the definition of cyber warfare, its moral underpinnings, and its effects on vital infrastructure.
\item To encourage more research in this field, we establish recommendations and make representations.
\end{itemize}

The format of this essay is as follows: The techniques used to choose the primary studies for analysis in a methodical manner are described in Section \ref{sec:2}. The results of all the primary research chosen are presented in Section \ref{sec:3}. The findings in relation to the earlier-presented study topics are discussed in Section \ref{sec:4}. The research is concluded in Section \ref{sec:5}, which also makes some recommendations for more study.

\section{Research Methodology}
\label{sec:2}
We conducted a systematic evaluation of several cyberwarfare research publications in order to reach the goal of answering the research questions. To prepare for our systematic literature review, we will evaluate the studies in-depth.

\subsection{Selection of primary studies}
\label{sec:2.1}
The search capability of a certain magazine or search engine was emphasized by entering keywords. The keywords were selected to support the growth of study results that would help in resolving the research's problems. On a number of websites, we searched for research papers and reviews utilizing keywords as "cyberwarfare," "cyberwar," "cybersecurity," "state-sponsored hacking," etc.

Passing keywords to a particular publication's or search engine's search function highlighted primary studies. The keywords were chosen to encourage the development of study findings that would aid in addressing the study's issues \cite{a4}. We looked for research articles and reviews using keywords like "cyberwarfare," "cyberwar," "cybersecurity," "state-sponsored hacking," etc. on a variety of web sites.

\vspace{1.5 mm}The platforms searched were:
\\- IEEE Xplore Digital Library 
\\- SpringerLink 
\\- Google Scholar
\\- UOG Library
\\- ScienceDirect
\\According to the search platforms, the title, keywords, or abstract were used in the searches. On October 15, 2022, we did the searches and processed all research that had submitted up to that point. The inclusion/exclusion criteria, which will be provided in Section \ref{sec:2.1}, were used to filter the results of these searches.

\subsection{Inclusion and Exclusion criteria}
Studies that are part of this systematic literature review should include empirical findings and may include case studies, hypothetical analyses, or studies that address the real consequences of cyberattacks organized by one country or organization against another. The requirements for admission are met by papers that discuss problems and difficulties in cyberwarfare. They must be written in English and subjected to peer review. Google Scholar might return papers that are of a lesser caliber, thus all results will be examined for conformity with these standards \cite{a10}. This Systematic Literature Review will only contain the most recent iteration of a study.

\subsection{Selection results}
The initial search results on the chosen platforms turned up a total of 61 studies. 52 were left after duplicate studies were eliminated. There were 20 publications left after the research were examined under the inclusion/exclusion criteria. The inclusion/exclusion criteria were applied again after reading all 20 articles, leaving 15 papers.

\subsection{ Quality assessment }
In accordance with the guidelines established by Kitchen ham and Charters \cite{kitchenham2007guidelines}, a review of the main studies' quality was conducted. This made it possible to evaluate the papers' applicability to the research topics while taking into account any indications of bias in the study and the reliability of the experimental results. The evaluation procedure was modeled after the method employed by Hosseini et al. \cite{Hosseini2019ASL}. Each and every one of the selected main studies is subjected to the following quality evaluation checklist.
\begin{table}[h!]
\caption{Inclusion and exclusion for primary studies} 
\label{Mytable2}
\setlength{\tabcolsep}{3pt}
\begin{tabular}{p{106pt} p{109pt} }
\hline
 Criteria for Inclusion & Criteria for exclusion\\
\hline
 \vspace{0.15 mm}The paper must present empirical data related
to cyberwarfare and hacking groups.  & \vspace{0.15 mm}Grey literature such as blogs and government documents\\
 \vspace{0.15 mm}The paper must contain information related to
cyberwarfare, cyber security, and hacking & \vspace{0.15 mm}Non-English papers \\
 \vspace{0.15 mm}The paper published in a conference
proceeding or journal & \vspace{0.15 mm} \\
\hline
\end{tabular}
\end{table}

The stages of finding the excluded studies, as seen in Table \ref{Mytable3}, are as follows:
\\Stage 1: \textbf{Cyberwarfare} The study of cyberwarfare, state-sponsored hacking, and information relating to cyber security must be the main topics of the paper.
\\Stage 2:\textbf{Context} The aims and conclusions of the research must be adequately contextualized. This will make it possible to understand the research correctly.
\\Stage 3: \textbf{Executing Cyberattack} The study must have sufficient information to accurately depict how the cyberattack is carried out against a particular infrastructure, which will help to address research questions RQ1 and RQ3.
\\Stage 4: \textbf{Ethical context} In an effort to help in answering RQ2, the paper must explain the morality of cyberwarfare.
\\Stage 5: \textbf{Data acquisition} To assess accuracy, specifics regarding the data's collection, measurement, and reporting must be provided.

\subsection{ Data extraction }

\begin{figure}[h]
	    \centering
	    \includegraphics[width=0.9\linewidth,keepaspectratio]{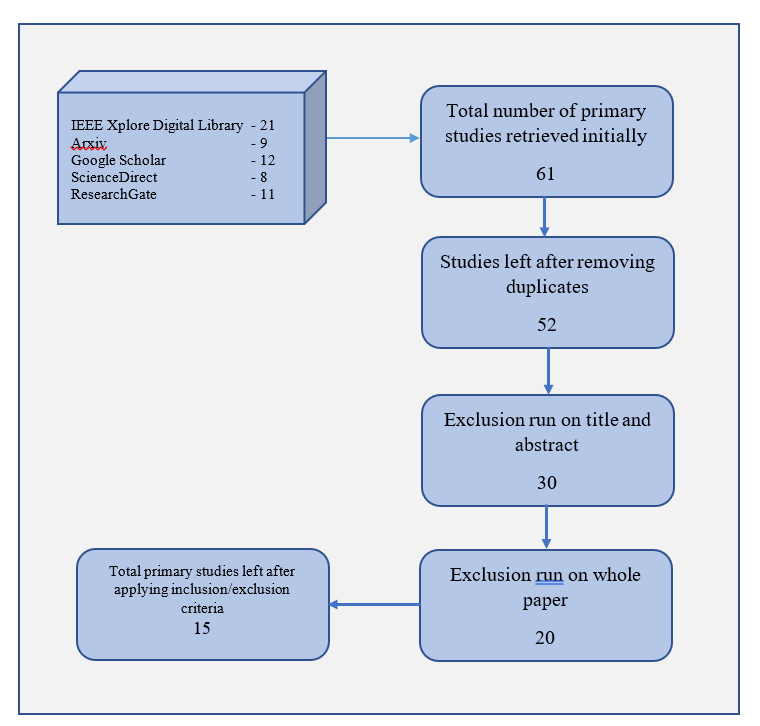}
	    \caption{Attrition of papers through processing}
	    \label{MyFig2}
\end{figure}

Data was then taken from all publications that had survived the quality evaluation in order to evaluate the completeness of the data and verify the accuracy of the information included within the articles. Before being applied to the whole set of research studies that have successfully completed the quality evaluation step, the data extraction technique was first tested on a sample of five studies. Each study's data was taken out, put into categories, and then entered into a spreadsheet. The data was categorized using the following categories:
\\Context data: Information on the study's objectives.
\\Qualitative data: Results and recommendations offered by the authors.
\\Quantitative data: Data obtained from testing and research are then applied to the study.
The attrition rate of papers obtained from the first keyword searches on each platform through to the final decision of primary studies are shown in Figure \ref{MyFig2} along with the quantity of articles chosen at each stage of the process.
\subsection{Data analysis}
We gathered the information included in the qualitative and quantitative data categories in order to achieve the goal of responding to the study questions. We also performed a meta-analysis on the studies that were exposed to the last step of data extraction.

\subsubsection{Publications over time}
\begin{figure}[h]
	    \centering
	    \includegraphics[width=0.9\linewidth,keepaspectratio]{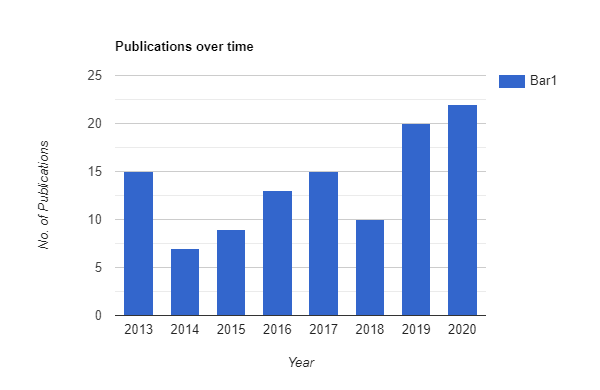}
	    \caption{publication over time}
	    \label{MyFig3}
\end{figure}

Despite being ubiquitous, the phrase "cyberwarfare" was less well-known due to its limited use.But after 2013, the public may access a sizable number of publications on cybersecurity and cyberwarfare. This may draw attention to how novel the concepts of cyberwarfare and cybersecurity are. The number of initial studies published each year is depicted in Figure \ref{MyFig3} as a chart. The employment of the terms "cybersecurity" and "cyberwarfare," as seen in the image, is on the rise. In the future, we anticipate seeing a rise in the number of publications related to cyberwarfare.

\subsubsection{Significant keyword counts:}
An analysis of keywords across all 42 studies was done in order to identify the recurring characteristics among the chosen primary research. The frequency of various key terms in all primary studies is shown in Table \ref{Mytable4}.
\begin{table}[h!]
\caption{Counts of the keywords in the primary studies.} 
\label{Mytable4}
\setlength{\tabcolsep}{5pt}
\begin{tabular}{p{125pt} p{90pt} }
\hline
\vspace{0.30 mm}Keywords   & \vspace{0.30 mm} Count\\
 \hline
 \vspace{-0.40 mm}cyberwarfare   & \vspace{-0.40 mm}1029\\
 \vspace{-1.50 mm}cyber security & \vspace{-1.50 mm} 526 \\
 \vspace{-1.50 mm}hacking & \vspace{-1.50 mm}120\\
 \vspace{-1.50 mm}critical infrastructure & \vspace{-1.50 mm} 254 \\
 \vspace{-1.50 mm}breach   & \vspace{-1.50 mm}20\\
 \vspace{-1.50 mm}information warfare & \vspace{-1.50 mm}  35  \\
 \vspace{-1.50 mm}cyber conflict & \vspace{-1.50 mm}92\\
 \vspace{-1.50 mm}cyberspace & \vspace{-1.50 mm} 74  \\
 \vspace{-1.50 mm}cyberattacks   & \vspace{-1.50 mm}141\\
 \vspace{-1.5 mm} denial-of-service & \vspace{-1.5 mm}  210  \\

 \hline
\end{tabular}
\end{table}
\section{Findings}
\label{sec:3}
\begin{figure}[h]
	    \centering
	    \includegraphics[width=0.9\linewidth,keepaspectratio]{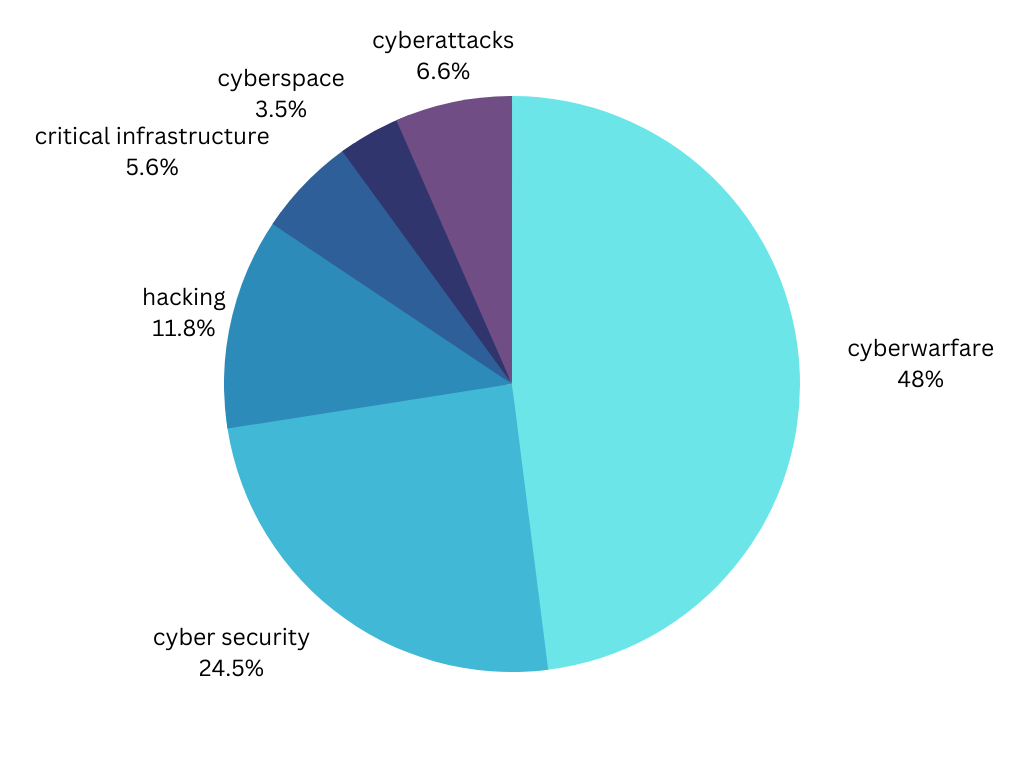}
	    \caption{Chart of themes of primary studies.}
	    \label{MyFig4}
\end{figure}

After carefully reading each main research article, pertinent qualitative and quantitative information was taken out and presented in Table \ref{Mytable5}. The main issue or emphasis of each of the key studies was how cyberwarfare is changing the globe. In Table \ref{Mytable5} below, the emphasis of each study is also listed.The percentages of the 30 main studies' various themes that allowed them to pass the quality evaluation and be incorporated into the data analysis are shown in Figure \ref{MyFig4}.

\begin{center}

\tablefirsthead{%
\hline
Primary Study & Key Qualitative \& Quantitative Data Reported & Theme\\
\hline}
\tablehead{%
\hline
Primary Study & Key Qualitative \& Quantitative Data Reported & Theme\\
\hline}
\tabletail{%
\hline}
\tablelasttail{\hline}
\bottomcaption{Main findings and themes of the primary studies}
\label{Mytable5}
\begin{mpsupertabular}{|p{1cm}|p{4.5cm}|p{2cm}|} % four columns, alignment for each
        
        \cite{s1} & {The fast expansion of the Internet has led to an expansion of cyberspace. Cyberwarfare has replaced traditional warfare as a result of the growth of cyberspace. Cyberspace has been recognized as the fifth battleground, after ground, sea, air, and space, particularly in the subject of defense.} & cyberspace \\
        
        \cite{s2} & {When compared to conventional combat, cyberwarfare is far cheaper and is conducted without endangering human life. Attacking a country or any personal data through cyberwarfare is on the rise nowadays.} & cyberwarfare \\
        
        \cite{https://doi.org/10.48550/arxiv.2208.10629} & {There has been a lot of discussion on how cyberattacks, hacktivists, and underground cybercrime have contributed to the war between Russia and Ukraine.} & cyberattacks \\

        \cite{s7} & {Many cyberattacks on critical systems in the past years will never be acknowledged owing to national security considerations.} & critical infrastructure \\
        
        \cite{s5} & {Due to the fast growth and widespread use of computer technology, cyberspace has evolved into an essential component of a government, a society, and people's daily lives.} & cyberspace \\
        
        \cite{s4} & {Cyberwarfare is the use of viral assaults or denial-of-service (DOS) attacks by any international body to target and make attempts to destroy the systems, infrastructure, or data management of another country.} & denial-of-service \\

        \cite{s6} & {Recent cyberattacks have blurred the distinction between real combatants and political hackers. This essay examines the implications of cybermilitias and the potential advantages and disadvantages of cyberconflict.} & hacking \\

        \cite{s8} & {The conventional tendency of cyber attacks against physical computer systems is toward controlling and command of the physical infrastructure.} & cyberattack \\
        
        \cite{s9} & {Human skills and capacities, including the capacity to guard, protect, assault, and win battles, must be tailored to the demands of cyberwarfare itself. A cyber unit must do all of these tasks in a variety of ways and in cooperation with critical institutions. These abilities represent the collective competency of a cyber troop.} & cyberwarfare \\
        
        \cite{s10} & {Because cyberwarfare differs from traditional kinetic warfare, it is important to examine fundamental warfare concepts in order to distinguish it from traditional armed conflict.} & cyberwarfare \\
        
        \cite{s11} & {In order to examine the mechanics and ramifications of offensive and defense cyberwarfare operations, Offensive and Protective Software Technologies is designed to bring along technical and non-technical cyberwarfare specialists, academics, and practitioners in relevant domains.} & Cyber security \\
        
        \cite{s12} & {Cybersecurity for vital infrastructure and services throughout the world is changing as a result of the hybrid conflict in Ukraine. For a very long time after the battle is resolved, the events will continue to alter cybersecurity practices all across the world.} & Cyber security \\
        
        \cite{s13} & {Cyberwarfare is the term used to describe politically inspired hacking for destruction and espionage. It is a sort of cyberwarfare that is occasionally contrasted with traditional warfare.} & cyberwarfare \\
        
        \cite{s14} & {Cyberwarfare entails operations that have a military or political purpose. Second, it entails using cyberspace to deliver straight or compounding kinetic impacts that produce outcomes equivalent to those of conventional military forces.Third, it produces outcomes that either lead to a major threat to a country's security, or that are taken in reaction to such a threat.} & cyberwarfare \\   
        \cite{s15} & {They can anticipate prolonged periods of low-intensity,global cyberwarfare, a Digital Cold War, while a play balance is pursued, and that cyberwarfare is not subject to control by international agreements.} & cyberwarfare \\
\end{mpsupertabular}

 \end{center}
Nearly half 48\% of all research are focused on cyberwarfare, according to the themes found in the primary studies. With a proportion of 24.5\% , cyber security is the second most widely discussed subject. With 11.8\% of the total themes, hacking is the third most prevalent and is largely concerned with how crucial political or financial information might be stolen by hackers. 

The fourth most frequent subject, with a 6.6\% share, is cyberattacks.Cyberspace makes up 3.5\% of primary study themes, whereas Critical Infrastructure makes up 5.6\%.
\section{Discussion}
\label{sec:4}
Initial search terms reveal that there are several papers that discuss cyberwarfare. Most of the chosen main studies are experimental hypotheses or notions with limited quantitative information and few real-world implications for challenges of today \cite{a7}."Cyberwar" and "cyberwarfare" are two different concepts. Cyberwarfare is the term used to describe the tactics, methods, and procedures that may be deployed in a cyberwar. The term "war" refers to a large-scale struggle that frequently lasts a long period and may include objectives including the use of force or the intention to kill.

Richard A. Clarke, who served as US President George W. Bush's chief cyber security advisor, described cyber war as "activities by a nation state to infiltrate another nation's systems and networks for the objective of causing injury or damage" (Knake and Clarke, 2010)\cite{borah2015cyber}.

According to the definition of cyberwar provided by the Oxford English Dictionary, it is only "another name for cyberwar." Cyber war, as described by the Oxford Dictionary (2013)\cite{oxford}, is "the use of digital technology to hamper a state or institution's functioning, notably the deliberate attack of communications infrastructure by some other country or organization." Like the other definitions under discussion, this one may be viewed as having issues.Like any activity that has the potential to hurt, cyberwarfare raises ethical issues. Particularly nations must comprehend if cyberwarfare is morally permissible and how to conduct it morally \cite{taddeo2012analysis}.All cyberattacks are currently orchestrated and planned by people. These tasks are projected to be created and performed in the future by artificial intelligence. Artificial intelligence (AI) technologies will be quicker than humans in analyzing and breaching security measures. They will have considerably greater ability to disrupt systems than they had in the past. One of the best methods to defend against cyberattacks will be blockchain. Systems can be kept safe, and data can be shielded from hackers.

Cyberwarfare is the useing of viral attacks or denial-of-service attacks by any international organization to target and try to destroy the computers, infrastructure, or data management of another country. Cyberwarfare is the use of pcs, IoTs, and networking in a conflict area or as a component of covert warfare. It involves aggressive and sceptical actions in respect to the risk of cyberattacks, monitoring, and collateral/other harm. Whether or not these actions fall within the definition of "war" is up for discussion. Transnational cyber-security is still a problem for major national and international entities.Current experts lack the means to deal with the ongoing change in the cybersecurity field since they lack in-depth knowledge of the issue. Both the culprit and the victim of these cyberattacks are mostly the United States. Following several cyberattacks, the nation has not implemented the required security to ensure against modern-day crime: cyberwarfare\cite{paper8}.

\subsubsection{RQ1: What is cyberwarfare?}
It is crucial to emphasize that the systematic literature evaluation under consideration will only concentrate on cyberattacks and warfare. In light of this, it should be mentioned that the researchers discovered a large number of papers on terrorism and conflict during the attrition process used to choose the primary studies. But the selection process gave special attention to research that have cyberwarfare at their core.

The term "cyber warfare," which is widely used in the establishment media, has a wide variety of connotations. Cyberwarfare, according to Alford's definition from 2001, is "any help to improve the overall to compel an adversary to carry out our national will, performed against the program controlling functions inside an adversary's system." \cite{AlfordCyberW}

This description from Alford represents the idea that nations would use cyberwarfare to further a national objective. However, it may be argued that modern combat does not necessarily strive to further such a cause. Possibly, the objective of modern warfare is to spread non-national philosophies and religious convictions. Therefore, it is imprudent to limit a definition of cyberwarfare to having that goal as its primary objective.

"Cyber warfare is the science and art of battling without fighting; of conquering an opponent without sacrificing their blood," says Jeffrey Carr in another explanation of the term \cite{paper10}. Unlike Alford's (2001), this term does not try to explain why the opposing sides are fighting. The idea that cyberwarfare won't result in casualties, however, has to be called into doubt. Losses of life might happen from a cyberattack on vital national infrastructure, including the electrical system.

\subsubsection{RQ2: Is cyberwarfare ethical?}
Like any act that has the potential to do harm, cyberwarfare poses ethical issues. Nations must precisely comprehend when and how to conduct such combat in cyberspace in accordance with moral principles. According to Taddeo, the Just War Theory, a set of precisely outlined criteria that define when a country is ethically justified in starting a war and how to act decently while at war, governs conventional battles. Taddeo argues that these ideas are difficult to apply to cyber warfare and that these difficulties necessitate further research\cite{Taddeo2012AnAF}.

Taddeo proposes three criteria that make up a "Just Cyber War" in an effort to solve the problem of cyber warfare ethics. These guidelines are connected to the concept of a "infosphere." This is described by Taddeo as "the setting in which informational things, both alive and lifeless, digital and analog, are ethically judged."

\begin{enumerate}
  \item Only entities that threaten or interfere with the infosphere's stability should be targeted by cyberwarfare.
  \item To protect the health of the infosphere, cyberwar should be fought.
  \item Fighting a cyberwar shouldn't be done to protect the infosphere.
\end{enumerate}

Point 1 stands for the idea that using cyberwar is acceptable to get rid of threats to the atmosphere's wellbeing. The next two arguments represent the idea that cyberwar should never be utilized to improve the well-being of the Infosphere beyond its natural condition, but rather solely to restore it to a status quo following a detrimental effect. Together, these arguments argue that cyberwar is morally acceptable so long as it helps to preserve the infosphere.

\subsubsection{RQ3: What kind of infrastructure is required to cope with cyberwarfare?}
Attacks by hackers on communication networks, governments, or power grids may bring entire civilizations to a standstill.As the battle in Ukraine shows, this makes them an effective military tool.The actions that can be taken to defend crucial infrastructure against cyberattacks:
\\
\textbf{Security Information and Event Management (SIEM):}safeguards against dangerous software and keeps track of network activities and access
\\
\textbf{Firewall:}Inbound communication between computer operations and the outside world is examined, evaluated, and filtered by a digital firewall.
\\
\textbf{Trust Zones:} Your internal network has extra firewalls created to safeguard sensitive information requiring increased protection.
\\
\textbf{Data Encryption:} Secure connections and data on personal devices are crucial for Iot network like smart metering and smart meters.
\\
\textbf{Multi-Factor Authentication:}requires staff to access a system or network using more than just a password

\section{Conclusion and future study}\label{sec:5}
This essay has offered an overview of current ideas about the difficulties posed by cyberwarfare. It started off by examining the terms cyberwar and cyberwarfare as they are now used. The United States should get ready for three crucial things, according to James Miller, a former assistant secretary of state of security for policy \cite{paper12}.
\begin{enumerate}
  \item Give strength for nuclear attack systems and for far-off conventional stages top priority, and invest in it.
  \item Make a concerted effort to project the nation's vital infrastructure and maintain a safety margin so that terrorist organizations are unable to put the nation at danger through cyberattacks.
  \item Create a type of script in advance and use it to direct and manage how you will react to a sizable cyberattack.
\end{enumerate}

The most important finding is that there is no one field that can adequately address all of the problems brought by cyberwarfare. For instance, attribution and cyber defense are undoubtedly technological concerns, but effective resolution of these and other problems need political, legal, and societal participation. Similar to this, developing a set of rules for cyber warfare calls for input from the technical and military on what would be possible to enforce in addition to legal experts.

\bibliographystyle{IEEEtran}
\bibliography{Ref}
\end{document}